# A Low Energy Beam Transport Design with high SCC for TAC Proton Accelerator

*A. Caliskan[1], H. F. Kisoglu[2], S. Sultansoy[3,4], M. Yilmaz[5]

[1]Department of Engineering Physics, Gumushane University, Gumushane, Turkey
[2]Department of Physics, Aksaray University, Aksaray, Turkey
[3]Institute of Physics, Academy of Sciences, Baku, Azerbaijan
[4]Physics Division, TOBB University of Economics and Technology, Ankara, Turkey
[5]Department of Physics, Gazi University, Ankara, Turkey

**Abstract**

In this study, a low energy beam transport (LEBT) channel for the proton linac section of the Turkic Accelerator Complex (TAC) has been designed by using TRACE 2D and TRAVEL codes. The LEBT channel is located between an ion source and a radio frequency quadrupole (RFQ) structure. The aims of the design studies are perfect matching between input and output beams with two solenoid magnets, small emittance growth and sufficient space for beam diagnostics. Total length of such LEBT channel is about 1.3 m. The current of $H^-$ ion beam from ion source is 80 mA. In the beam dynamical simulations, we have taken into account some space charge compensation (SCC) factors between %93.75 and %100. The results of both codes have been compared for the selected SCC factors. Additionally, beam aperture study for % 95 SCC factor has been done to discard unwanted particles and increase the beam brightness.

## 1. Introduction

The Turkic Accelerator Complex (TAC) is a regional project [1] and has been developed with support of the Turkish State Planning Organization (DPT) by collaboration of 10 Turkish universities. Its conceptual design report was completed in 2005 and technical design report (TDR) studies have been continued since 2006. Today, TAC project includes linac-ring type super charm factory, synchrotron light source based on positron ring, free electron laser based on electron linac, GeV scale proton accelerator and TAC test facility [2].

The proton accelerator construction will have 3 MeV, 100 MeV and 1 GeV phases. It will give an opportunity to produce secondary muon and neutron beams for applied research fields in additional to primary proton beam. In the muon region, a lot of applied investigations in some research areas like high-Tc superconductivity, phase transitions, impurities in semiconductors will be performed using powerful muon spin resonance (μSR) method [3]. In the neutron region, it is planned to use for different fields of science such as engineering, molecular biology and fundamental physics. Additionally, accelerator driven systems (ADS) technology application of GeV energy proton accelerator becomes very attractive for our country since Turkey has essential thorium reserves [4].

*Corresponding author: acaliskan@gumushane.edu.tr

The proton accelerator is linear and its fundamental accelerator structures are ion source, low energy beam transport (LEBT) channel, RFQ (radio-frequency quadrupole) and medium energy beam transport (MEBT) channel for the low energy section. The low energy beam transport channel is located between an ion source and a RFQ. Its task is transfering and matching the ion beam to the RFQ by using solenoid magnets. Forming into bunched structure of the beam is realized in RFQ structure while the beam obtained from ion source has un-bunched structure. Acceleration, focusing and bunching of the ion beam are simultaneously performed in RFQ by using only transverse electric fields. Axial longitudinal electric fields, needed for beam acceleration process, are produced by internal surface modulation of each electrode inside RFQ. There is no acceleration in the first section of RFQ structure called "radial matching section". After the beam formed into bunched structure, acceleration starts. For medium energy section, we have planned using of drift tube linac (DTL) [5, 6] and coupled cavity drift tube linac (CCDTL) cavities. For high energy section, using of normal conducting or super conducting cavities is under investigation.

## 2. Low Energy Beam Transport Line

One of the possible reasons of the emittance growth and beam halo formation is space charge effects, caused by space charge forces, dominate the beam at low energies. Space charge compensation method has been experimentally using to decrease the space charge effects in the beam inside the LEBT channel [7]. In this method, the ion beam passes through the residual gas inside the beam line. Ionization of the molecules of this residual gas has been performed as a result of the interaction between ion beam and gas molecules. The secondary particles, produced by ionization process such as electrons or ions, are trapped until a steady state reached. This situation in the beam can be considered as plasma that creates a focusing effect which compensates the space charge effects [8].

TRAVEL [9] beam dynamics simulation code, which includes space-charge effects of a continuous beam, and the TRACE 2D [10] code have been used to design such a LEBT line and study beam dynamics. In the simulation studies with both codes, we have taken into account the effect of the space charge compensation method in beam dynamics. $H^-$ beam extracted from an ion source was matched with the beam parameters at RFQ plane during these simulations. Thus, this study provides a basis for RFQ. The parameters of such a LEBT line intended for TAC proton accelerator are given in Table 1.

Table 1 LEBT line parameters

| Parameters | Length (mm) |
|---|---|
| **Drift** | 50 |
| **Solenoid** | 300 |
| **Drift** | 500 |
| **Solenoid** | 300 |
| **Drift** | 150 |
| **Beam aperture ($r_0$)** | 50 |

*Corresponding author: acaliskan@gumushane.edu.tr

As is seen above, the LEBT line is planned to consist of two solenoids and three drifts. Aperture size in which the beam goes through was chosen as 5 cm along the whole LEBT line as first stage. In this way, we can see general behavior of the beam in case of no particle loss.

## 3. Beam Dynamics Simulations

It is foreseen that the H⁻ beam extracted from the source has 80 mA peak current and 75 keV kinetic energy at the entrance of the LEBT. Such a beam, which has initial parameters of $\varepsilon_{xx'} = \varepsilon_{yy'} = 0.20$ mm.mrad (norm. rms emittance), $\beta_x = \beta_y = 11$ mm/mrad and $\alpha_x = \alpha_y = -2$, was passed through the LEBT line with different SCC factors. Figure 1 shows the entrance beam profiles. In beam dynamics, SCC factor of % 100 is equivalent to zero-current (no space charge effect) simulation. This is ideal situation of space charge compensation technique. To leave an error margin of %6.25 from ideal situation, we have scanned the some SCC factor from % 93.75 (5 mA) to % 100 (0 mA).

Magnetic fields of the solenoids have been firstly determined by the TRAVEL code simulations and entrance beam of LEBT has been matched with the beam at RFQ plane, which has parameters of $\beta_x = \beta_y = 0.1$ mm/mrad and $\alpha_x = \alpha_y = 1.5$ (shown in Figure 2), for each SCC factors. Also, as is seen from Table 2, mismatch factors calculated by TRAVEL are below 0.056 which means perfect matching. Mismatch factors are the same for both *x* and *y* planes as the beam is symmetrical. Mismatch factor theoretically given as;

$$M = \left[\frac{1}{2}\left(R + \sqrt{R^2 - 4}\right)\right]^{1/2} - 1$$

where $R = \beta_1\gamma_2 + \beta_2\gamma_1 - 2\alpha_1\alpha_2$. The $\alpha_1, \beta_1, \gamma_1$ and $\alpha_2, \beta_2, \gamma_2$ are Twiss parameters of initial and final beam respectively [10].

*Corresponding author: acaliskan@gumushane.edu.tr

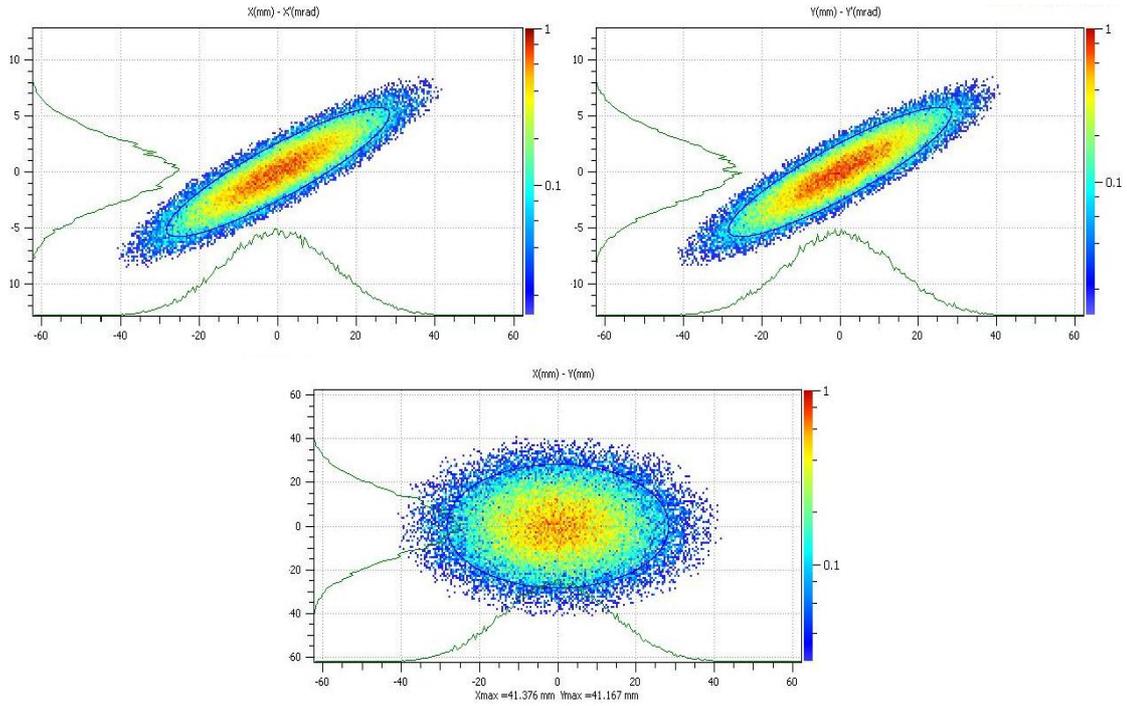

**Figure 1** Beam profiles at the entrance of the LEBT line.

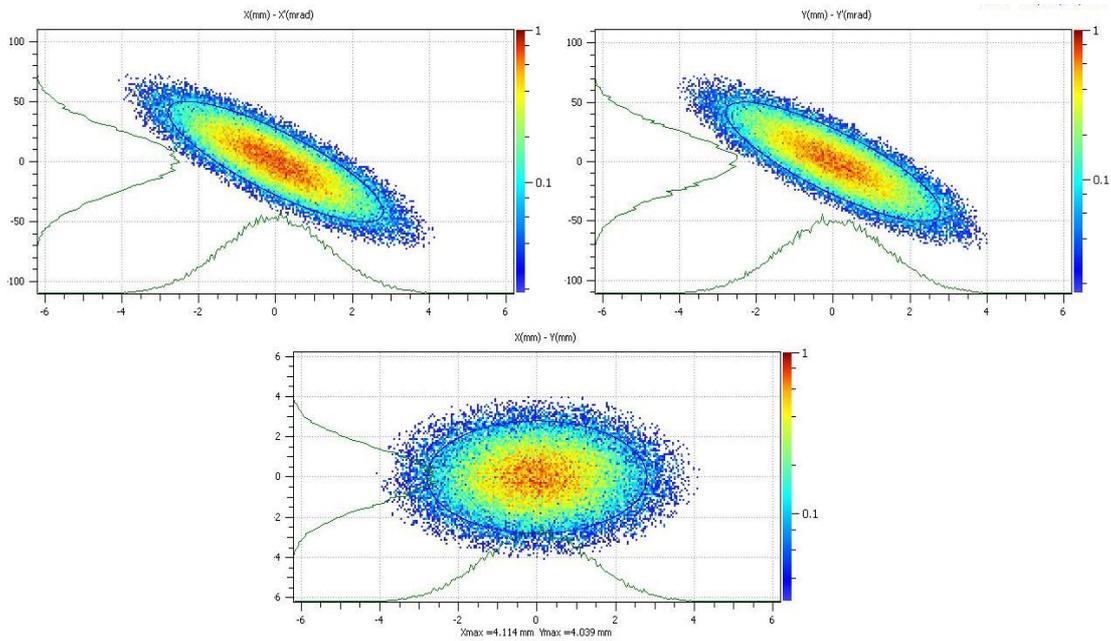

**Figure 2** The beam profiles at the RFQ plane.

In the second stage TRACE 2D was run for the same magnetic fields, obtained from TRAVEL, to get mismatch factors in comparison with TRACE 2D. All results of simulations are given in Table 2.

*Corresponding author: acaliskan@gumushane.edu.tr

Table 2 Magnetic fields of solenoids and mismatch factors for both codes

| Beam Current (mA) | Magnetic Field (T) | | Mismatch Factor (TRAVEL) | Mismatch Factor (TRACE 2D) |
|---|---|---|---|---|
| | Solenoid-1 | Solenoid-2 | | |
| 0 | 0,223 | 0,435 | 0,001 | 0,033 |
| 0,5 | 0,121 | 0,220 | 0,010 | 0,245 |
| 1 | 0,228 | 0,432 | 0,010 | 0,351 |
| 1,5 | 0,230 | 0,430 | 0,026 | 0,667 |
| 2 | 0,161 | 0,899 | 0,023 | 0,661 |
| 2,5 | 0,114 | 0,245 | 0,025 | 1,02 |
| 3 | 0,112 | 0,250 | 0,056 | 1,17 |
| 3,5 | 0,114 | 0,250 | 0,031 | 1,46 |
| 4 | 0,242 | 0,421 | 0,023 | 2,04 |
| 4,5 | 0,244 | 0,420 | 0,003 | 2,32 |
| 5 | 0,249 | 0,413 | 0,018 | 2,52 |

The results of both codes are compatible only for zero-current according to Table 2. At the existence of space charge forces, mismatch factors of TRACE 2D simulations are bigger than ones of TRAVEL simulations. It is clearly seen that mismatch factor and beam current are directly proportional. The reason of inconsistency of the results of both codes is based on the difference at the program algorithms. TRACE 2D is an envelope tracking code and its emittance is constant while TRAVEL is a Particle-in Cell (PIC) code. TRACE 2D matches two different beam ellipses (*x* and *y* planes) and its calculation algorithm more different from ones of the TRAVEL code at the existence of space charge forces. Matched beam ellipses of LEBT line for % 100 space charge compensation obtained from TRACE 2D can be seen in Figure 3.

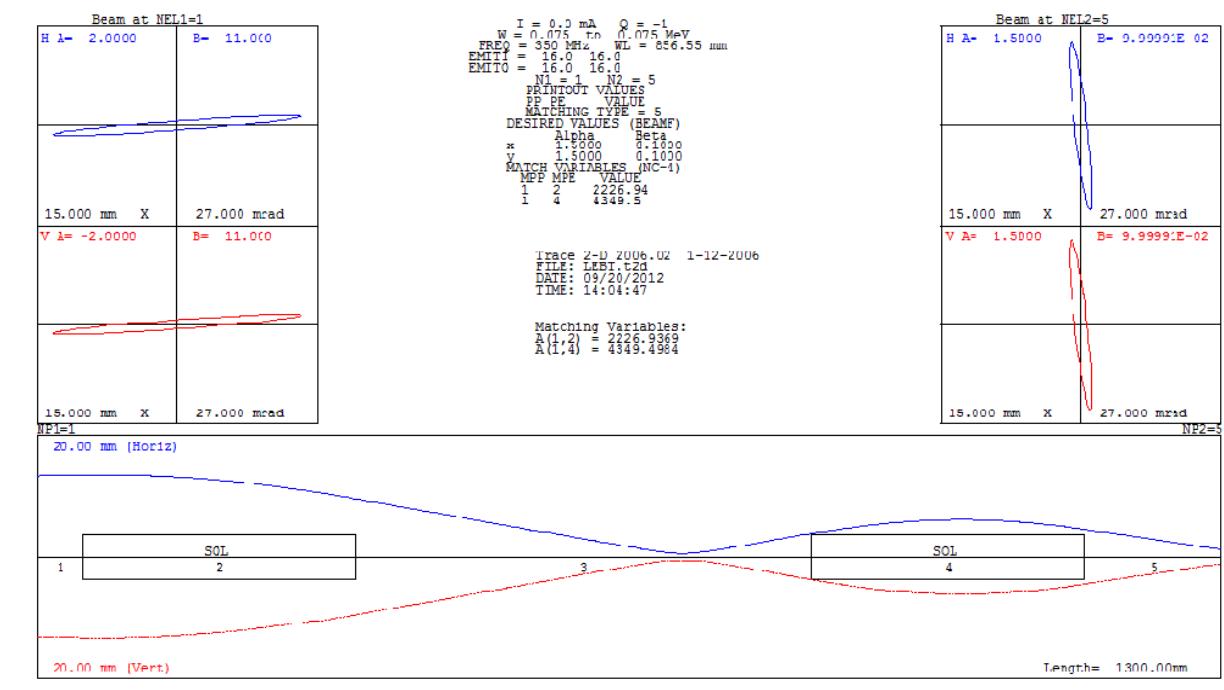

Figure 3. Beam matching result from TRACE 2D for % 100 space charge compensation

*Corresponding author: acaliskan@gumushane.edu.tr

## 4. The Collimator Study for % 95 SCC factor

Beam aperture at drifts (it is constant at solenoids) in the LEBT line can be used as collimator. Thus, we can avoid unwanted particles which cause to increase of emittance and size of the beam by using collimator at drift regions in the LEBT line. Also brightness of the beam can be increase in this way. So, the beam can be transferred to next parts intensively. Nevertheless if we use collimator decreasing aperture size, particle losses will be higher as well. Thusly, changing of alive particles in the beam versus variation of aperture size has been obtained as Figure 4.

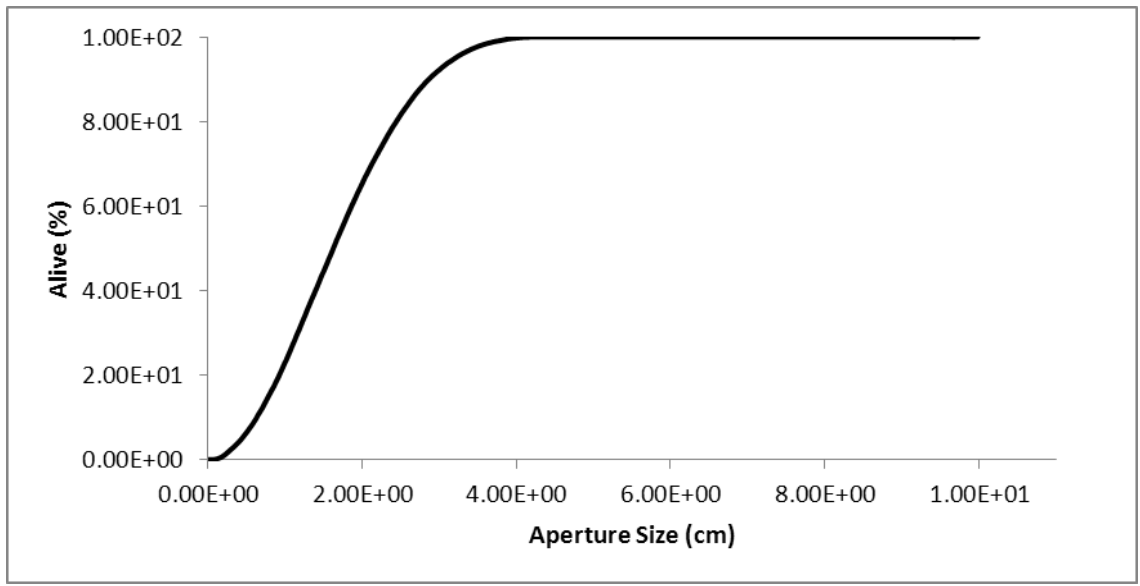

**Figure 4** Variation of beam transmission vs. aperture size of LEBT line.

We can see from Figure 4 that there is no particle loss at 5 cm, roughly. If we minimize the aperture size transmission of the beam is decreased and the cost of the LEBT getting be more less since the cost of LEBT is proportional to $r_0^2$ where $r_0$ is aperture size [11]. We chose the aperture size as 3,21 cm for usage of collimator. For this value transmission of the beam is %95 and *M* mismatch factor is 0.06 roughly. Aperture size could be taken smaller for decreasing of cost of LEBT. But the smaller aperture size we choose, the more particle losses we get. Also *M* mismatch factor will be higher, if we minimize it. The simulation results for this aperture size value are shown in Figure 5 and Figure 6. As is seen, the beam size is %6,74 averagely and normalized rms emittance is %14,12 smaller than the case with no collimator. So that, brightness of the beam is %10,73 higher as a result of decreasing of the emittance. The exit beam distribution for both situations (with/without collimator) can be seen in Figure 7. In this Figure, the beam losses of % 5 can be seen clearly. The two cases with no collimator and with collimator for %95 SCC are summarized in Table 3. Consequently, the beam is still matched to RFQ input plane and it is more stable in spite of the beam losses of % 5 with usage of collimator.

*Corresponding author: acaliskan@gumushane.edu.tr

Table 3 Comparing of the cases with collimator and without collimator for %95 SCC.

| | Transmission (%) | Norm. RMS Emittance (m.rad) | Brightness (mA/mm.mrad) |
|---|---|---|---|
| **No collimator** | 100 | $2.96 \times 10^{-7}$ | $13.51 \times 10^{6}$ |
| **3.21 cm collimator** | 95 | $2.54 \times 10^{-7}$ | $14.96 \times 10^{6}$ |

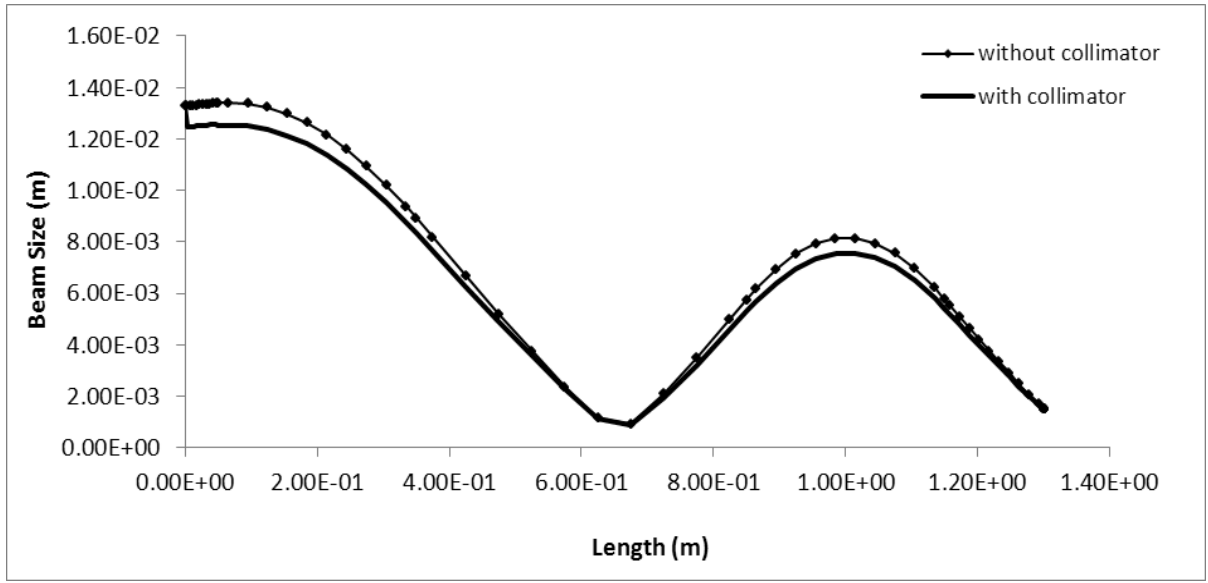

Figure 5. Beam size changing along the LEBT

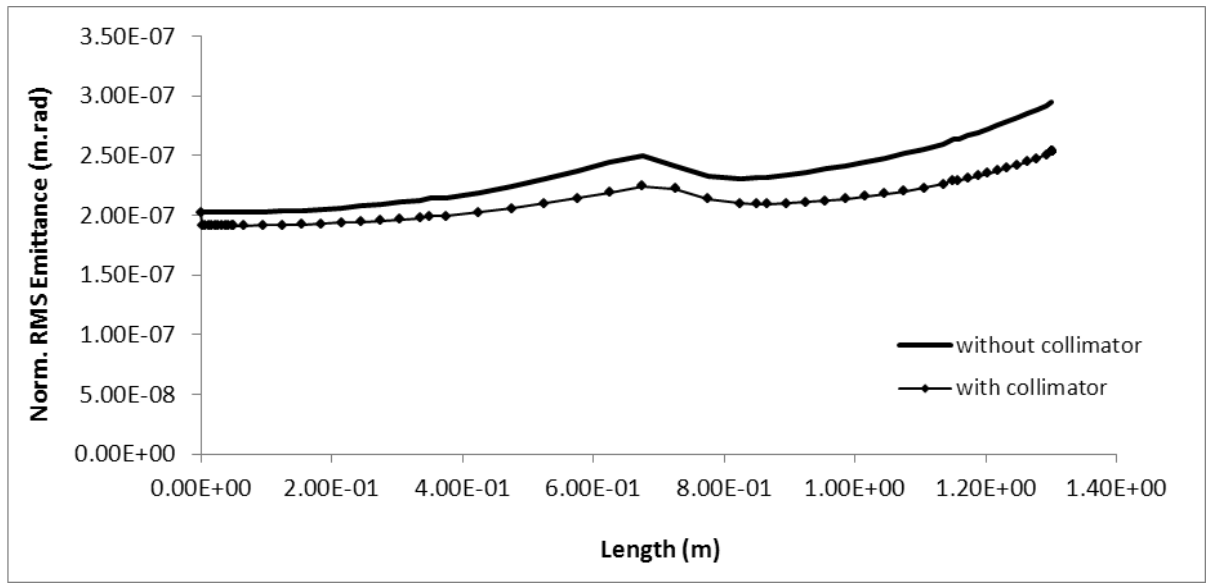

Figure 6. Normalized rms emittance evolution throughout the line

*Corresponding author: acaliskan@gumushane.edu.tr

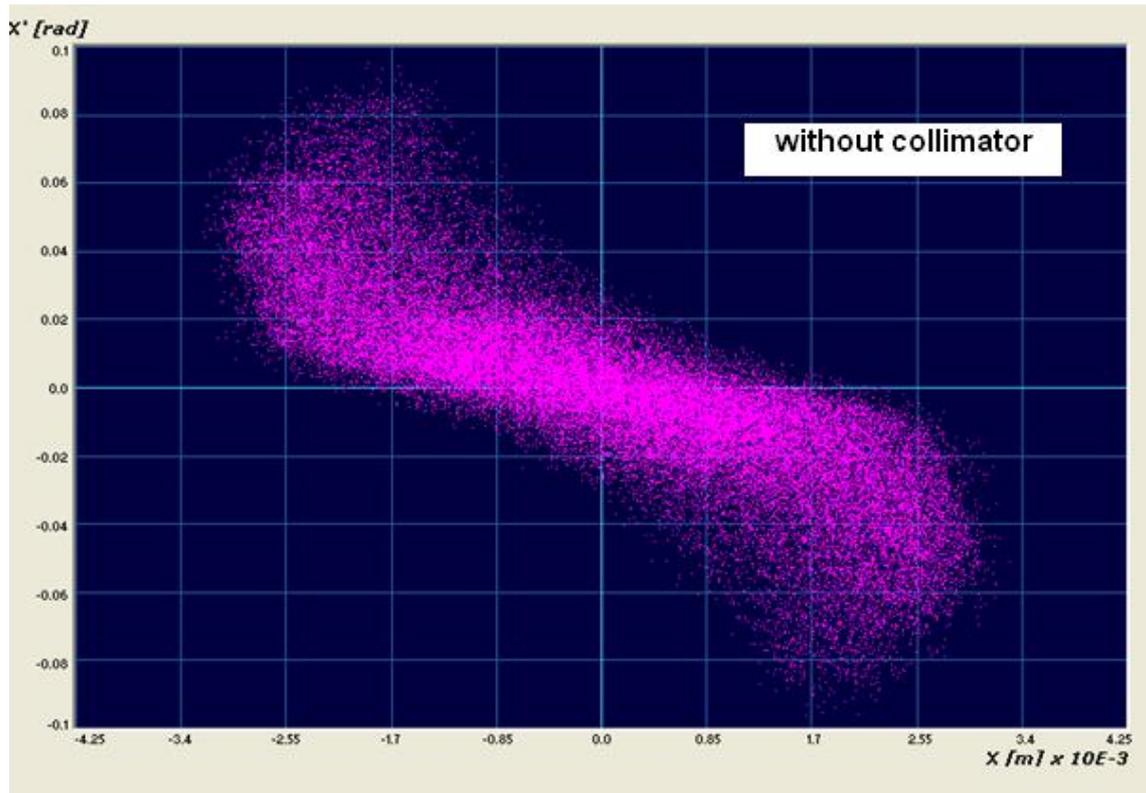

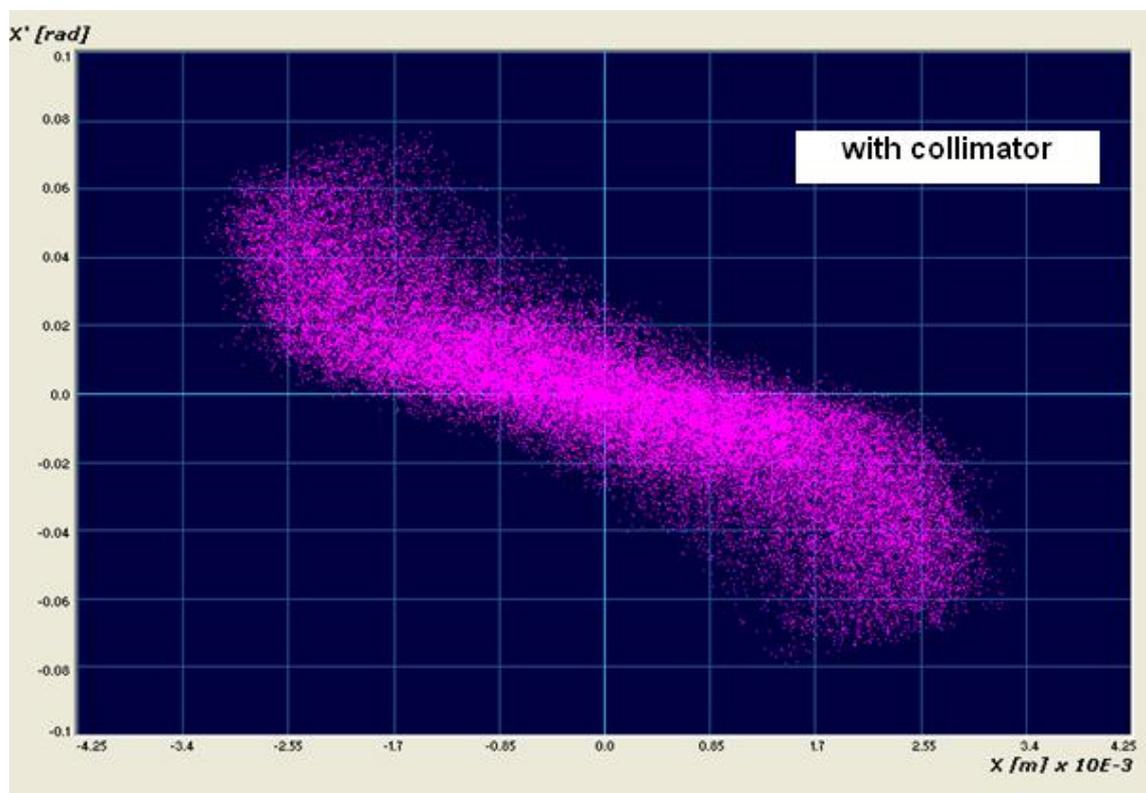

Figure 7. A comparision of the final particle distribution in phase space with/without collimator

*Corresponding author: acaliskan@gumushane.edu.tr

## 5. Conclusions

Before this study, accelerator parts (DTL, CCDTL, CCL) of proton linac of the TAC project has been designed [5, 6, 12]. The design studies of the pre-accelerator parts are ongoing. Thus, we have done this study to shed light especially on RFQ. For this purpose a low energy beam transport channel has been designed. H$^-$ ion beam extracted from source with 80 mA peak current and 75 keV kinetic energy matched with RFQ parameters by tuning the magnetic fields of solenoids. In the beam dynamics simulations, TRAVEL and TRACE 2D codes have been used by taking into account the some SCC factors from % 93.75 to % 100. For each value in this interval, magnetic fields of solenoids have been determined by TRAVEL simulation code. When the results of two codes are compared it is seen that there are essential differences under the influence of the space charge forces because of the program algorithms. TRACE 2D gives totally different results as it is a beam envelope code while TRAVEL is a PIC mode code. So, according to our results TRAVEL is more reliable than TRACE 2D in beam dynamics simulations.

Then, we have performed a study to determine the ideal beam aperture for % 95 SCC factor to see effects of usage of collimator on the beam. Taking into account beam losses, mismatch factor and cost of LEBT the aperture size has been chosen as 3.21 cm. Matching of the beam and essential enhancement of the brightness, although small fraction of the beam has been lost, is achieved using collimator.

## Acknowledgements

We would like to thank Turkish State Planning Organization (DPT), under the grants no DPT-2006K120470, and Scientific and Technological Research Council of Turkey (TUBITAK) for supports.

## References

[1] Sultansoy, S., Regional project for elementary particle physics: linac-ring type c-τ-factory, Turk. J. Phys. 17, 591-597 (1993).

[2] Turkic Accelerator Complex web page: http://thm.ankara.edu.tr

[3] Caliskan, A., Proton synchrotron and muon region applications, Master Thesis, Science and Technology Institute, Gazi University, Ankara, Turkey (2005).

[4] Arik, M., Bilgin, P.S., Caliskan, A., Cetiner, M. A., Sultansoy, S., A provisional study of ADS within Turkic Accelerator Complex project, III. Internetional Conference on Nuclear & Renewable Energy Resources (NuRER), 20-23 May, Istanbul, Turkey (2012).

[5] Caliskan, A., Yilmaz M., DTL cavity design and beam dynamics for a TAC linear proton accelerator, Chinese Physics C, 36(2):167-172 (2012).

[6] Caliskan, A., Kisoglu, H.F., Yilmaz, M., Some criteria at DTL design for Turkish Accelerator Center (TAC) linac, Azerbaijan Journal of Physics, XVI (2):534-536 (2010).

*Corresponding author: acaliskan@gumushane.edu.tr

[7] Ferdinand, R., et al., Space-charge neutralization measurement of a 75-keV 130-mA hydrogen-ion beam, Proceedings of the Particle Accelerator Conference (PAC97), 2723-2725, 12-16 May, Vancouver, Canada(1997).

[8] Chauvin, O., et al., Transport of intense ion beams and space charge compensation issues in low energy beam lines, Review of Scientific Instruments, 02B320:83 (2012).

[9] Perrin, A., et al., "*TRAVEL v4.07 User Manual*", CERN, April 2007.

[10] Crandall, K.R. and Rusthoi, D.P., "*Trace 3-D Documentation*", Los Alamos National Laboratory, LA-UR-97-886(1997).

[11] Baartman, R., "*Low energy beam transport design optimization for RIBs*", NIM B Beam Interactions with Materials & Atoms, 10.1016/S0168-583X(02)02103-1.

[12] Caliskan, A., Parameter Optimization and Beam Dynamics for 1 GeV Energy and Multipurpose Linear Proton Accelerator, PhD Thesis, Science and Technology Institute, Gazi University, Ankara, Turkey (2011).

*Corresponding author: acaliskan@gumushane.edu.tr